\documentclass[a4paper,11pt]{article}
\pdfoutput=1 

\usepackage{jcappub} 

\usepackage[T1]{fontenc} 
\usepackage[utf8]{inputenc}

\title{\boldmath Modified Starobinsky inflation by the $R\ln\left(  \square\right)  R$ term}

\author{J. Bezerra-Sobrinho}
\author{and L. G. Medeiros}
\affiliation{Escola de Ci\^{e}ncias e Tecnologia, 	Universidade Federal do Rio Grande do Norte,\\
	Campus Universit\'{a}rio, s/n - Lagoa Nova, CEP 59072-970, Natal, Rio Grande do Norte, Brazil}

\emailAdd{jeremias.bs@gmail.com}
\emailAdd{leo.medeiros@ufrn.br}

\abstract{In the context of effective theories of gravity, a minimalist bottom-up
approach which takes into account $1$-loop quantum corrections leads to
modifications in the Einstein-Hilbert action through the inclusion of four
extra terms: $R^{2}$, $C_{\kappa\rho\alpha\beta}C^{\kappa\rho\alpha\beta}$,
$R\ln\left(  \square\right)  R$ and $C_{\kappa\rho\alpha\beta}\ln\left(
\square\right)  C^{\kappa\rho\alpha\beta}$. The first two terms are necessary
to guarantee the renormalizability of the gravitational theory, and the last
two terms (nonlocal terms) arise from the integration of massless/light matter
fields. This work aims to analyze how one of the nonlocal terms, namely
$R\ln\left(  \square\right)  R$, affects the Starobinsky inflation. We
consider the nonlocal term as a small correction to the $R^{2}$ term, and we
demonstrate that the model behaves like a local model in this context. In
addition, we show that the approximate model in the Einstein frame is
described by a canonical scalar field minimally coupled to general relativity.
Finally, we study the inflationary regime of this model and constrain its free
parameters through observations of CMB anisotropies.}

\begin{document}
\maketitle
\flushbottom

\section{Introduction}

The inflationary period is defined as an accelerated expansion, usually almost
exponential, in the pre-nucleosynthesis universe. The central goals of
inflation are to solve the flatness and horizon problems and mainly to
generate the inhomogeneities that provide the initial conditions for the
structure formation \cite{Guth1981,Linde1983,MukBook}.

There is a large number of inflationary models in the literature
\cite{Enciclopedia2014,BamMyrOdiSeb2014,MyrOdiSeb2014,EliOdiSebMyr2017,KosModRacSta2016,KosKumSta2017,KosKumMazSta2020,KosKumSta2020}.
These models can be classified based on their common
properties, such as the variability of their fields -- e.g. small and large
fields inflation \cite{Brand2016} -- or the number of free parameters they
possess \cite{Enciclopedia2014}. Complex models tend to have more parameters,
and they usually better fit the observations. On the other hand, the
introduction of extra degrees of freedom decreases the predictability power of
the model. In this sense, the most desirable is a model with a smaller number
of free parameters that satisfies current observations
\cite{Planck2018,Bicep2021}. Another important guide in building an
inflationary model is its theoretical foundation. Conceptually, well-motivated
models generated by extensions of general relativity or the standard model of
particle physics are more relevant than their purely phenomenological counterparts.

To satisfy the three aspects pointed out in the previous paragraph --
consistency with the observations, few parameters, and theoretically
well-grounded -- is a non-trivial task. Nevertheless, we can cite a few
examples such as Higgs inflation \cite{BezShapo2008}\ and Starobinsky model
\cite{Sta1980} which fulfill these criteria.

The Higgs inflation is an inflationary model whose standard Higgs scalar field
is non-minimally coupled to gravity through $\xi\left\vert h\right\vert ^{2}R$
term \cite{BezShapo2008}. This model has only one free parameter, perfectly
satisfying the cosmological CMB observations. Furthermore, from a theoretical
point of view, the model is well justified since the $\xi\left\vert
h\right\vert ^{2}R$ term is necessary for the renormalizability of scalar
fields in curved spacetimes \cite{CaCoJa1970}. Despite its original success,
the Higgs inflationary model presents some issues such as the generation of
large quantum corrections for $\xi>>1$ \cite{BurLeeTro2009,Hert2010}%
\footnote{The large quantum corrections arise for any energy scale bigger than
$M_{P}/\xi$.} and the possibility of triggering Higgs field vacuum decay
\cite{Rubio2019,MarRaSto2019}.

The Starobinsky model is an inflationary model of modified gravity where an
$R^{2}$ term is included in the Einstein-Hilbert action \cite{Sta1980}. Like
the Higgs Inflation, the Starobinsky model properly describes current
cosmological observations from a single free parameter. In addition,
Starobinsky inflation provides clear predictions for observables such as the
scalar spectral index and the tensor-to-scalar ratio
\cite{Planck2018,Bicep2021}.

From a theoretical point of view, Starobinsky inflation is based on a
bottom-up approach of quantum gravity. In the context of effective theories
and taking into account up to $1$-loop quantum corrections, the action for the
effective quantum gravity can be written as
\cite{BaVi1985,BuOdSh1992,DonMen2014,Mag2016,Tei2020}
\begin{equation}
S=\frac{M_{P}^{2}}{2}\int d^{4}x\sqrt{-g}\left[  R+\frac{1}{2\kappa_{0}}%
R^{2}+\frac{1}{2\kappa_{2}}C^{2}+\mathcal{L}_{NL}\right]\text{,} \label{eff action}%
\end{equation}
where $\kappa_{0}$ and $\kappa_{2}$ are dimensional constants, $C^{2}$ is the
Weyl invariant, i.e. $C^{2}=C_{\mu\nu\alpha\beta}C^{\mu\nu\alpha\beta}$ and
$\mathcal{L}_{NL}$ contains the gravitational corrections which arise from the
integration of matter fields.

Structurally, the term $C^{2}$ has the same importance as $R^{2}$ since both
have the fourth mass dimension and are necessary to guarantee the $1$-loop
renormalizability of the theory \cite{HoVe1974,Stelle1977}. A difficulty in
dealing with the $C^{2}$ term is that it generates ghost-like fields, and the
quantization of this type of field is no longer trivial \cite{PaUh1950}. Among
the techniques used to quantize ghost fields we can mention the introduction
of an undefined metric in Hilbert space \cite{LeeWick1969,SalStru2015} and the
use of PT-antilinear symmetry \cite{BenMan2007,BenMan2008}. Even though these
techniques allow a consistent quantization process, they may generate problems
in the probabilistic interpretation of the theory due to the loss of
unitarity. One way to deal with this problem is to consider that ghost fields
are unstable, and therefore they do not contribute to the asymptotic spectrum
of the theory \cite{ModSha2016,DoMe2019,Anselmi2017,AnsPiv2018}. Another
possibility is to define a non-trivial norm between states of ghost fields
in order to recover the unitarity of the theory \cite{Salvio2019,Salvio2021}. In the
inflationary context, these issues show up during the quantum process of
generating primordial fluctuations. By taking into account the aspects
mentioned above, recent works explore the influence of the Weyl invariant on
inflation and show how it affects the tensor-to-scalar ratio
\cite{IvaTok2016,Salvio2017,AnBiPi2020}.

For the inflationary period is connected to a hot Big-Bang universe (via
reheating \cite{KoLinSta1997,BaTsuWan2006,Loz2019}), matter fields must be
present, even though they are negligible during the inflationary regime. In
the context of effective theories, the presence of these fields gives rise to
non-trivial gravitational corrections that are encapsulated in the $\mathcal{L}_{NL}$
term. The form of this term is complicated and depends on the relationship
between the energy scale adopted and the masses of the matter fields
\cite{Mag2016}. Considering the energy scale as the inflationary scale and
assuming fields with masses far below this value\footnote{This is exactly the
case for the particles of standard model.}, the term $\mathcal{L}_{NL}$ gets a nonlocal
structure which in the bilinear curvature approximation is described by
\cite{DonIvaShke2017}
\begin{equation}
\mathcal{L}_{NL}=\frac{2\alpha}{M_{P}^{2}}R\ln\left(  \frac{\square}{\mu^{2}%
}\right)  R+\frac{2\beta}{M_{P}^{2}}C_{\kappa\rho\alpha\beta}\ln\left(
\frac{\square}{\bar{\mu}^{2}}\right)  C^{\kappa\rho\alpha\beta}\text{,}
\label{Non local Lagrangian}
\end{equation}
where $\mu$ and $\bar{\mu}$ coefficients are the renormalization points. The
dimensionless constants $\alpha$ and $\beta$ are not free parameters, and they
can be calculated from the effective action which takes into account the
$1$-loop quantum correction generated by the integration of massless/light
fields \cite{Mag2016,Tei2020}. The specific values of $\alpha$ and $\beta$
depend on the number of matter fields and their respective spins
\cite{DonMen2014,CaCaKu2019}. Furthermore, due to the non-minimum coupling of
scalar fields with scalar curvature via $\xi\phi^{2}R$ term, the parameter
$\alpha$ also depends on $\xi$.

The discussion presented in the previous three paragraphs provides a natural
theoretical framework in which the Starobisnky inflation is embedded. Thus, it
is reasonable to expect the terms $C^{2}$, $R\ln\left(  \square\right)  R$ and
$C_{\kappa\rho\alpha\beta}\ln\left(  \square\right)  C^{\kappa\rho\alpha\beta
}$ can generate corrections to the Starobinsky model. Our paper aims to
explore the effects of one of these terms, namely $R\ln\left(  \square\right)
R$, on Starobinsky inflation. The study will be carried out considering that
the nonlocal term can be treated analytically and provides small corrections
to the Starobinsky model. In this situation, we will show that our model can
be rewritten in a local form whose dynamics are described by a single scalar field.

The manuscript is organized as follows. The nonlocal gravitational model and
its field equations in the Jordan Frame are presented in Section
\ref{sec - Acao nao local}. The perturbative approach used to deal with the
nonlocal term and the transition to the Einstein frame are developed in Section
\ref{sec - perturbative}. In Section \ref{sec - inflation}, the description of
the inflationary regime is performed and the model's free parameters are
constrained. The final comments are presented in Section
\ref{sec - final comments}.

\section{Nonlocal gravitational action\label{sec - Acao nao local}}

We start by considering an effective gravitational action which differs from
Starobinsky action by the $R\ln\left(  \square\right)  R$ term:

\begin{equation}
S=\frac{M_{P}^{2}}{2}\int d^{4}x\sqrt{-g}\left[  R+\frac{1}{2\kappa_{0}}%
R^{2}+\frac{2\alpha}{M_{P}^{2}}R\ln\left(  \frac{\square}{\mu^{2}}\right)
R\right]\text{,} \label{acao S}%
\end{equation}
where $\kappa_{0}$ is a positive free parameter with squared mass units, $\mu$
is the renormalization point, and $\alpha$ is a dimensionless parameter that
depends on the light matter fields present in the fundamental theory. We
choose as renormalization point the inflation energy scale ($E_{\inf}%
\sim\kappa_{0}^{1/2}$) and consider the parameter $\alpha$ as a free
parameter.\footnote{In the context of effective theories, the parameter
$\alpha$ is fixed only in the case we know the number of matter scalar fields
and the intensity of their respective non-minimum couplings with the scalar
curvature \cite{DonMen2014}.} In addition, it will be assumed that the
nonlocal operator $\ln\left(  \square\right)  R$ has an analytical
representation around the adopted energy scale. Thus,%
\begin{equation}
\ln\left(  \frac{\square}{\mu^{2}}\right)  R =\sum_{n=1}^{\infty}%
\frac{\left(  -1\right)  ^{n-1}}{n}\left(  \frac{\square}{\mu^{2}}-1\right)
^{n}R
=\sum_{n=1}^{\infty}\sum_{k=0}^{n}Z_{k,n}\left(  \frac{\square}{\mu^{2}%
}\right)  ^{k}R\text{,} \label{rep. serie 2}%
\end{equation}
where
\begin{equation}
Z_{k,n}\equiv\frac{\left(  -1\right)  ^{k+1}(n-1)!}{k!\left(  n-k\right)
!}\text{.} \label{Z_kn}%
\end{equation}
It is worth mentioning that there are several papers in the literature that
use analytical representations to deal with nonlocal operators present in
higher-order modified gravity \cite{BiMaSi2006,BiKoMa2010,BiGeKoMa2012,BiKoMa2013,TaBiMa2015}.

An important point to be discussed is the validity of the analytic
representation for the nonlocal operator. We know the series
(\ref{rep. serie 2}) converges only when
\begin{equation}
0<x<2\text{ \ \ \ where \ \ \ }x=\frac{\square}{\mu^{2}}\text{.} \label{convergence}%
\end{equation}
In principle, this restriction seems to limit the feasibility of Eq.
(\ref{rep. serie 2}). However, we are interested in describing the
inflationary regime, and in this period, the energy remains approximately
constant.\footnote{The invariant
$
R=6\left(  \frac{\ddot{a}}{a}+\frac{\dot{a}^{2}}{a^{2}}\right)
$
is a very slow varying function during inflation. Thus, terms of type $\square^{k}R$ remain
approximately constant.} Thus, by choosing as renormalization point $E_{\inf}%
$, we guarantee that the representation (\ref{rep. serie 2}) remains valid
throughout all inflation. Also, note that due to the lower limit of Eq.
(\ref{convergence}), the series representation remains valid after inflation,
even though the convergence speed decreases as the system moves away from
$E_{\inf}$. It should also be emphasized that the choice to represent the
operator $\ln\left(  \square\right)  R$ in terms of a series neglects
non-analytical effects, which could be described by an integral representation
\cite{EsMuVa2005,CaEs2007,DonMen2014}. This choice is justified because, in
the scope of effective theories, the physical effects regarded are always
within a well-defined energy range. In the specific case of the action
(\ref{acao S}), this range is located below the Planck scale and (far) above
the masses of the matter fields.

By introducing convenient Lagrange multipliers and using the equations of
motion, we can rewrite the action (\ref{acao S}) in the Jordan frame (see
appendix \ref{ApJD}). Defining the dimensionless scalar fields%
\begin{align}
\lambda &  \equiv\frac{R}{\kappa_{0}},\label{lambda}\\
\theta &  \equiv1+\lambda+b\ln\left(  \frac{\square}{\mu^{2}}\right)
\lambda \text{,}\label{theta}%
\end{align}
we get
\begin{equation}
S  = \frac{M_{P}^{2}}{2}\int d^{4}x\sqrt{-g}\left\{  \theta R+\kappa
_{0}\left(  1-\theta\right)  \lambda
 +\frac{\kappa_{0}}{2}\lambda^{2}+\frac{\kappa_{0}b}{2}\lambda
\ln\left(  \frac{\square}{\mu^{2}}\right)  \lambda\right\}\text{,}
\label{acao frame Jordan lambda}%
\end{equation}
where $b\equiv4\alpha\kappa_{0}/M_{P}^{2}$ is a dimensionless parameter that
represents the effectiveness of the nonlocal term concerning the Starobinsky term.

The sign of parameter $b$ depends on the value of $\alpha$ since $\kappa_{0}$
is a strictly positive quantity. Negative values of $\alpha$ are physically
more relevant because they correspond to values obtained from effective
theories. Considering the action (\ref{Non local Lagrangian}) and taking into
account the standard model matter fields,\footnote{We are neglecting the
graviton.} we get \cite{DonMen2014,CaCaKu2019}
\begin{equation}
\alpha=-\frac{5\left(  6\xi-1\right)  ^{2}}{11520\pi^{2}}N_{s}%
\text{,} \label{alpha definition}%
\end{equation}
where $N_{s}=4$ takes into account the internal degrees of freedom of the
Higgs field, and $\xi$ is the coupling constant present in the term
$\xi\left\vert h\right\vert ^{2}R$.\footnote{The difference in sign between
Eq. (\ref{alpha definition}) and the result of Ref. \cite{CaCaKu2019} comes
from the distinct definitions associated with the nonlocal Lagrangian terms.}
It is also worth noting that spinorial and vector contributions are null in
the Weyl-Weyl basis \cite{DonMen2014}. Thus, in the approach of effective
theories, $\alpha$ is always negative.

Another indication that negative $b$ is a more physically consistent choice
comes from the approximation of the $\ln\square$ series by its first term. By
carrying out this approximation, we obtain%
\begin{equation}
\frac{2\alpha}{M_{P}^{2}}R\ln\left(  \frac{\square}{\mu^{2}}\right)
R\simeq\frac{2\alpha}{M_{P}^{2}}R\left(  \frac{\square}{\mu^{2}}-1\right)  R\text{.}
\label{App}
\end{equation}
The contribution of the $R\square R$ term was studied in the inflationary
\cite{CuMePo2018} and weak field \cite{SilMed2020} contexts, and in both
cases, it was shown that for $\alpha>0$, the system is affected by instabilities.

The above arguments indicate that negative $b$ is physically more relevant.
However, as these arguments are not definitive, we will consider both signs
for the value of $b$.

\subsection{Field equations in Jordan frame\label{sec - Field Equations}}

Let's determine the field equations associated with the action
(\ref{acao frame Jordan lambda}). The first step is to rewrite the nonlocal
term $\lambda\ln\left(  \square\right)\lambda$ in terms of a series in the form
(\ref{rep. serie 2}):
\begin{align}
S_{NL}  = &\frac{M_{P}^{2}}{2}\int d^{4}x\sqrt{-g}\frac{\kappa_{0}b}{2}%
\lambda\ln\left(  \frac{\square}{\mu^{2}}\right)  \lambda\nonumber\\
= &\frac{M_{P}^{2}\kappa_{0}b}{4}\sum_{n=1}^{\infty}\sum_{k=0}^{n}%
\frac{\left(  -1\right)  ^{n-1}}{n}\frac{\left(  -1\right)  ^{n-k}%
n!}{k!\left(  n-k\right)  !}\frac{1}{\mu^{2k}}
\int d^{4}x\sqrt{-g}\lambda\left(  \square^{k}\lambda\right)
\text{.}\label{Termo nao local}%
\end{align}
Thus, the action (\ref{acao frame Jordan lambda}) becomes
\begin{equation}
S=\frac{M_{P}^{2}}{2}\int d^{4}x\sqrt{-g}\left[  \theta R+\kappa_{0}\left(
1-\theta\right)  \lambda+\frac{\kappa_{0}}{2}\lambda^{2}\right]
+S_{NL}\text{ .}\label{acao em partes}
\end{equation}
By taking the variation of $S$ concerning $g^{\mu\nu}$, we get
the field equation
\begin{align}
\theta\left(  R_{\mu\nu}-\frac{1}{2}g_{\mu\nu}R\right)& -\left(
\nabla_{\mu}\nabla_{\nu}\theta\right)  +g_{\mu\nu}\left[  \left(
\square\theta\right)  -\frac{\kappa_{0}}{2}\left(  1-\theta\right)
\lambda-\frac{\kappa_{0}}{4}\lambda^{2}\right] \nonumber\\
& -\frac{\kappa_{0}b}{4}g_{\mu\nu}\lambda\ln\left(  \frac{\square}{\mu^{2}
}\right)  \lambda +\frac{\kappa_{0}b}{4}\sum_{n=1}^{\infty}\sum_{k=1}^{n}
\frac{\left(  -1\right)  ^{n-1}}{n}\frac{\left(  -1\right)  ^{n-k}
n!}{k!\left(  n-k\right)  !}
{\displaystyle\sum\limits_{l=1}^{k}}
P_{\mu\nu}^{k} =0\text{,}\label{Field1}
\end{align}
where%
\begin{align}
P_{\mu\nu}^{k}  =&
{\displaystyle\sum\limits_{l=1}^{k}}
\left\{  g_{\mu\nu}\left(  \frac{\square}{\mu^{2}}\right)  ^{l-1}%
\lambda\left(  \frac{\square}{\mu^{2}}\right)  ^{k-l+1}\lambda+g_{\mu\nu
}\left[  \frac{\nabla_{\rho}}{\mu}\left(  \frac{\square}{\mu^{2}}\right)
^{l-1}\right]  \lambda\left[  \frac{\nabla^{\rho}}{\mu}\left(  \frac{\square
}{\mu^{2}}\right)  ^{k-l}\right]  \lambda\right.  \nonumber\\
&  -\left.  2\left[  \frac{\nabla_{\mu}}{\mu}\left(  \frac{\square}{\mu^{2}%
}\right)  ^{l-1}\right]  \lambda\left[  \frac{\nabla_{v}}{\mu}\left(
\frac{\square}{\mu^{2}}\right)  ^{k-l}\right]  \lambda\right\}
\text{.}\label{Pmunuk frame Jordan}%
\end{align}
See appendix \ref{Ap 3} for details. Furthermore, substituting Eq. (\ref{theta}) in Eq. (\ref{Field1}) we can
rewrite the equation of metric as
\begin{align}
\theta \left(  R_{\mu\nu}-\frac{1}{2}g_{\mu\nu}R\right) & -\nabla_{\mu}%
\nabla_{\nu}\theta+g_{\mu\nu}\left[  \square\theta-\frac{\kappa_{0}}{4}\left(
1-\theta\right)  \lambda\right]  \nonumber\\
& +\frac{\kappa_{0}b}{4}\sum_{n=1}^{\infty}\sum_{k=1}^{n}\frac{\left(
-1\right)  ^{n-1}}{n}\frac{\left(  -1\right)  ^{n-k}n!}{k!\left(  n-k\right)
!}P_{\mu\nu}^{k} =0\text{.}\label{Eq. de campo Jordan 3}%
\end{align}

Finally, a dynamic equation for the $\theta$ field can be obtained from the
trace of Eq. (\ref{Eq. de campo Jordan 3}) and the relation (\ref{lambda}):%
\begin{equation}
3\square\theta-\kappa_{0}\lambda+\frac{\kappa_{0}b}{4}\sum_{n=1}^{\infty}%
\sum_{k=1}^{n}\frac{\left(  -1\right)  ^{n-1}}{n}\frac{\left(  -1\right)
^{n-k}n!}{k!\left(  n-k\right)  !}P^{k}=0\text{,}\label{Eq. de campo Jordan 5}%
\end{equation}
where $P^{k}$ is the trace of $P_{\mu\nu}^{k}$ given by
\begin{equation}
P^{k}  =
{\displaystyle\sum\limits_{l=1}^{k}}
\left\{  4\left(  \frac{\square}{\mu^{2}}\right)  ^{l-1}\lambda\left(
\frac{\square}{\mu^{2}}\right)  ^{k-l+1}\lambda
+2\left[  \frac{\nabla_{\rho}}{\mu}\left(  \frac{\square}{\mu^{2}%
}\right)  ^{l-1}\right]  \lambda\left[  \frac{\nabla^{\rho}}{\mu}\left(
\frac{\square}{\mu^{2}}\right)  ^{k-l}\right]  \lambda\right\}
\text{.}\label{Pk frame Jordan}%
\end{equation}

The equations (\ref{Eq. de campo Jordan 3}), (\ref{Eq. de campo Jordan 5}) and
(\ref{theta}) are the dynamic equations for the fields $g^{\mu\nu}$, $\theta$
and $\lambda$. In addition, in the limit of $b\rightarrow0$, we recover the
equations from the Starobinsky model in the Jordan frame:%
\begin{align*}
3\square\theta-\kappa_{0}\lambda &  =0\text{,}\\
\theta\left(  R_{\mu\nu}-\frac{g_{\mu\nu}}{2}R\right)  -\nabla_{\mu}%
\nabla_{\nu}\theta+g_{\mu\nu}\left[  \square\theta+\frac{\kappa_{0}}{4}%
\lambda^{2}\right]   &  =0\text{,}
\end{align*}
where $\lambda=\theta-1.$

\section{Perturbative approach\label{sec - perturbative}}

The presence of the nonlocal term makes the field equations obtained in Section
\ref{sec - Field Equations} quite complicated. Because of it, we will develop
a perturbative approach considering the nonlocal part, regulated by parameter
$b$, as a small correction to the Starobinsky model. In this case, we will
only consider first-order corrections on $b$.

In zero-order the Eqs. (\ref{Eq. de campo Jordan 5}) and (\ref{theta}) result
in%
\begin{equation}
\square\lambda=\frac{\kappa_{0}}{3}\lambda\text{ \ \ or \ \ }%
\square\theta=\frac{\kappa_{0}}{3}\left(  \theta-1\right)
.\label{ordem zero theta e lambda}%
\end{equation}
So, by induction, we get\footnote{The approach used below is similar to that developed in Ref. \cite{BiMaSi2006}.}
\begin{equation}
\square^{n}\lambda=\left(  \frac{\kappa_{0}}{3}\right)  ^{n}\lambda
\Rightarrow\square^{n}\lambda=\left(  \frac{\kappa_{0}}{3}\right)
^{n-1}\square\lambda.\label{Ansatz_lambda1}%
\end{equation}

All nonlocal terms are at least first order terms. Thus, we can use eq.
(\ref{Ansatz_lambda1}) to simplify them. Let's start with the nonlocal term of
the equation (\ref{theta}):%
\begin{align}
\ln\left(  \frac{\square}{\mu^{2}}\right)  \lambda &  =\sum_{n=1}^{\infty}%
\sum_{k=0}^{n}\frac{\left(  -1\right)  ^{n-1}}{n}\frac{\left(  -1\right)
^{n-k}n!}{k!\left(  n-k\right)  !\mu^{2k}}\square^{k}\lambda\nonumber\\
& = \sum_{n=1}^{\infty}\sum_{k=0}^{n}\frac{\left(  -1\right)  ^{n-1}}{n}%
\frac{\left(  -1\right)  ^{n-k}n!}{k!\left(  n-k\right)  !\mu^{2k}}\left(
\frac{\kappa_{0}}{3}\right)  ^{k-1}\square\lambda
=\frac{3}{\kappa_{0}}
\ln\left(  \frac{\kappa_{0}}{3\mu^{2}}\right)  \square\theta
.\label{Aprox termo Eq Theta}%
\end{align}
Note that between the first and second lines, we use the right-hand version of
Eq. (\ref{Ansatz_lambda1}). It is justified because we want to preserve a
differential structure associated with the scalar fields. In addition, by
keeping only second-order derivatives, we obtain the simplest possible
differential form for the fields $\lambda$ and $\theta$. It is also important
to stress that the series representation used is valid only if%
\begin{equation}
\mu^{2}>\frac{\kappa_{0}}%
{6}.\label{Raio de convergencia}%
\end{equation}
The above expression determines the convergence radius of the series
representations and constrains the choice of the renormalization point.

Similar calculations for the equations (\ref{Pmunuk frame Jordan}) and
(\ref{Pk frame Jordan}) result in%
\begin{align}
P_{\mu\nu}^{k} & =
{\displaystyle\sum\limits_{l=1}^{k}}
\left\{  g_{\mu\nu}\left(  \frac{\kappa_{0}}{3\mu^{2}}\right)  ^{l-1}%
\lambda\left(  \frac{\kappa_{0}}{3\mu^{2}}\right)  ^{k-l}\frac{\square\lambda
}{\mu^{2}}
+  g_{\mu\nu}\left[  \frac{\nabla_{\rho}}{\mu}\left(  \frac
{\kappa_{0}}{3\mu^{2}}\right)  ^{l-1}\right]  \lambda\left[  \frac
{\nabla^{\rho}}{\mu}\left(  \frac{\kappa_{0}}{3\mu^{2}}\right)  ^{k-l}\right]
\lambda\right.  \nonumber\\
&  -\left.  2\left[  \frac{\nabla_{\mu}}{\mu}\left(  \frac{\kappa_{0}}%
{3\mu^{2}}\right)  ^{l-1}\right]  \lambda\left[  \frac{\nabla_{v}}{\mu}\left(
\frac{\kappa_{0}}{3\mu^{2}}\right)  ^{k-l}\right]  \lambda\right\}\Rightarrow
\nonumber
\end{align}
\begin{equation}
P_{\mu\nu}^{k} =\frac{3}{\kappa_{0}}\left(  \frac{\kappa_{0}}{3\mu^{2}}\right)
^{k}k\left[  g_{\mu\nu}\left(  \lambda\square\lambda+\nabla_{\rho}%
\lambda\nabla^{\rho}\lambda\right)  -2\nabla_{\mu}\lambda\nabla_{v}%
\lambda\right]  ,\label{Aprox Pmunuk}%
\end{equation}
and%
\[
P^{k} =
{\displaystyle\sum\limits_{l=1}^{k}}
\left\{  4\left(  \frac{\kappa_{0}}{3\mu^{2}}\right)  ^{l-1}\lambda\left(
\frac{\kappa_{0}}{3\mu^{2}}\right)  ^{k-l}\frac{\square\lambda}{\mu^{2}%
}
+  2\left[  \frac{\nabla_{\rho}}{\mu}\left(  \frac{\kappa_{0}}%
{3\mu^{2}}\right)  ^{l-1}\right]  \lambda\left[  \frac{\nabla^{\rho}}{\mu
}\left(  \frac{\kappa_{0}}{3\mu^{2}}\right)  ^{k-l}\right]  \lambda\right\}\Rightarrow
\]
\begin{equation}
P^{k} =k\frac{6}{\kappa_{0}}\left(  \frac{\kappa_{0}}{3\mu^{2}}\right)
^{k}\left[  2\lambda\square\lambda+\nabla_{\rho}\lambda\nabla^{\rho}%
\lambda\right]  .\label{Aprox Pk}
\end{equation}
Substituting these last two expressions into Eqs. (\ref{Eq. de campo Jordan 3})
and (\ref{Eq. de campo Jordan 5}) and performing the sums, we obtain the
field equations in their approximate form:%
\begin{align}
\theta \left(  R_{\mu\nu}-\frac{1}{2}g_{\mu\nu}R\right) & -\nabla_{\mu}%
\nabla_{\nu}\theta+g_{\mu\nu}\left[  \square\theta-\frac{\kappa_{0}}{4}\left(
1-\theta\right)  \lambda\right]  \nonumber\\
& +\frac{3b}{4}\left[  g_{\mu\nu}\left(  \lambda\square\lambda+\nabla_{\rho
}\lambda\nabla^{\rho}\lambda\right)  -2\nabla_{\mu}\lambda\nabla_{v}%
\lambda\right] \simeq0,\label{Eq Jordan local 1}%
\end{align}
and
\begin{equation}
3\square\theta-\kappa_{0}\lambda+\frac{3b}{2}\left[  2\lambda\square
\lambda+\nabla_{\rho}\lambda\nabla^{\rho}\lambda\right]  \simeq
0,\label{Eq Jordan local 2}%
\end{equation}
and%
\begin{equation}
\lambda+1-\theta+\frac{3b}{\kappa_{0}}\ln\left(  \frac{\kappa_{0}}{3\mu^{2}%
}\right)  \square\theta\simeq0.\label{Eq Jordan local 3}%
\end{equation}

Finally, we can use Eq. (\ref{Eq Jordan local 3}) to substitute $\lambda$ in
the equations (\ref{Eq Jordan local 1}) and (\ref{Eq Jordan local 2}). By
keeping only first-order corrections we get

\begin{align}
\theta\left(  R_{\mu\nu}-\frac{1}{2}g_{\mu\nu}R\right) & -\left(  1+\frac
{3b}{2}\right)  \nabla_{\mu}\theta\nabla_{v}\theta \nonumber \\ & + g_{\mu\nu}\left\{
\square\theta+\frac{\kappa_{0}}{4}\left(  \theta-1\right)  ^{2}+\frac{3b}%
{4}\left[  K\left(  \theta-1\right)  \square\theta+\nabla_{\rho}\theta
\nabla^{\rho}\theta\right]  \right\} \simeq 0,\label{Eq de campo aprox Jordan 1}
\end{align}
and
\begin{equation}
\left[  1+b\left(  \theta-K\right)  \right]  \square\theta-\frac{\kappa_{0}%
}{3}\left(  \theta-1\right)  +\frac{b}{2}\nabla_{\rho}\theta\nabla^{\rho
}\theta \simeq 0,\label{Eq de campo aprox Jordan 2}
\end{equation}
where $K=1-\ln\left(  \kappa_{0}/3\mu^{2}\right)  $. Note that the perturbative
approach allows writing the field equations as a
set of local differential equations for the fields $g_{\mu\nu}$ and $\theta$.

\subsection{Einstein frame\label{sec - Frame Einstein}}

In order to simplify the subsequent analysis, let's rewrite the field
equations (\ref{Eq de campo aprox Jordan 1}) and
(\ref{Eq de campo aprox Jordan 2}) in the Einstein frame. Performing the
transformations \cite{NoOdOi2017,CuMeMePo2019}%
\[
\theta=e^{\chi}\text{ \ and \ }g_{\mu\nu}=e^{-\chi}\bar{g}_{\mu\nu
}\text{ \ }\Rightarrow\text{ \ }R_{\mu\nu}=\bar{R}_{\mu\nu}+\nabla_{\mu}%
\nabla_{\nu}\chi-\frac{1}{2}\nabla_{\mu}\chi\nabla_{\nu}\chi+\frac{1}{2}%
g_{\mu\nu}\left(  \square\chi+\nabla^{\beta}\chi\nabla_{\beta}\chi\right)  ,
\]
we obtain
\begin{align}
\bar{R}_{\mu\nu}-\frac{1}{2}\bar{g}_{\mu\nu}\bar{R} & -\frac{3}{2}\left(
1+be^{\chi}\right)  \left[  \bar{\nabla}_{\mu}\chi\bar{\nabla}_{\nu}\chi
-\frac{1}{2}\bar{g}_{\mu\nu}\bar{\nabla}^{\beta}\chi\bar{\nabla}_{\beta}%
\chi\right] \nonumber \\ & +\frac{3b}{4}\bar{g}_{\mu\nu}\left(  e^{\chi}-1\right)
K\bar{\square}\chi+\frac{\kappa_{0}}{4}\bar{g}_{\mu\nu}\left(  1-e^{-\chi
}\right)  ^{2} \simeq0,\label{Eq campo aprox Eins 1}
\end{align}
and
\begin{equation}
\left(  1+be^{\chi}\right)  \bar{\square}\chi-bK\bar{\square}\chi+\frac{1}%
{2}be^{\chi}\bar{\nabla}_{\rho}\chi\bar{\nabla}^{\rho}\chi-\frac{\kappa_{0}%
}{3}e^{-\chi}\left(  1-e^{-\chi}\right) \simeq 0.\label{Eq campo aprox Eins 2}%
\end{equation}
Note that by performing the transition to the Einstein frame only after the perturbative
treatment, we avoid having to deal with conformal transformations in the $R\square^{k} R$ operator.

In these two expressions, we see that the nonlocal corrections appear in two
different ways: $b$ alone and $be^{\chi}$. In addition, for $e^{\chi}>>1$,
which usually occurs during the inflationary regime, it is possible the term
$be^{\chi}$ is not small even if $b<<1$. This observation shows that the
linear approximation should not be performed in terms containing $be^{\chi}$.

The next step is to apply the perturbative approach to deal with the terms
$\bar{\square}\chi$ present in the Eqs. (\ref{Eq campo aprox Eins 1}) and
(\ref{Eq campo aprox Eins 2}). In zero-order the equation
(\ref{Eq campo aprox Eins 2}) is given by%
\[
\bar{\square}\chi\simeq\frac{\kappa_{0}}{3}e^{-\chi}\left(  1-e^{-\chi
}\right)  .
\]
Replacing this result in the Eqs. (\ref{Eq campo aprox Eins 1}) and
(\ref{Eq campo aprox Eins 2}) we get
\begin{equation}
\bar{R}_{\mu\nu} -\frac{1}{2}\bar{g}_{\mu\nu}\bar{R}-\frac{3}{2}\left(
1+be^{\chi}\right)  \left[  \bar{\nabla}_{\mu}\chi\bar{\nabla}_{\nu}\chi
-\bar{g}_{\mu\nu}\frac{1}{2}\bar{\nabla}^{\beta}\chi\bar{\nabla}_{\beta}
\chi\right] +\frac{\kappa_{\alpha}}{4}\bar{g}_{\mu\nu}\left(  1-e^{-\chi}\right)  ^{2}
\simeq0,\label{Eq field Einstein 1}%
\end{equation}
and%
\begin{equation}
\left(  1+be^{\chi}\right)  \bar{\square}\chi+\frac{1}{2}be^{\chi}\bar{\nabla
}_{\rho}\chi\bar{\nabla}^{\rho}\chi-\frac{\kappa_{\alpha}}{3}e^{-\chi}\left(
1-e^{-\chi}\right)  \simeq0,\label{Eq field Einstein 2}%
\end{equation}
where%
\begin{equation}
\kappa_{\alpha}\equiv\kappa_{0}\left(  1+bK\right)  .\label{kappa alpha}%
\end{equation}

Lastly, we can redefine the scalar field $\chi$ to obtain a canonical kinetic
term. By carrying out the change \cite{CaSoShaSta2018}%
\begin{equation}
\bar{\partial}_{\mu}\chi=\frac{d\chi}{d\phi}\bar{\partial}_{\mu}\phi\text{
\ \ \ where \ \ \ }\frac{d\chi}{d\phi}=\frac{\sqrt{2}}{M_{P}\sqrt{3\left(
1+be^{\chi}\right)  }},\label{d chi / d psi}%
\end{equation}
we get%
\[
\bar{\square}\chi=\frac{d\chi}{d\phi}\bar{\square}\phi-\frac{3M_{P}^{2}}%
{4}be^{\chi}\left(  \frac{d\chi}{d\phi}\right)  ^{4}\bar{\partial}^{\rho}%
\phi\bar{\partial}_{\rho}\phi,
\]
and the Eqs. (\ref{Eq field Einstein 1}) and (\ref{Eq field Einstein 2}) are
rewritten as%
\begin{equation}
\bar{R}_{\mu\nu} -\frac{1}{2}\bar{g}_{\mu\nu}\bar{R}-\frac{1}{M_{P}^{2}%
}\left[  \bar{\partial}_{\mu}\phi\bar{\partial}_{\nu}\phi-\frac{1}{2}\bar
{g}_{\mu\nu}\bar{\partial}^{\beta}\phi\bar{\partial}_{\beta}\phi\right]
+\frac{\kappa_{\alpha}}{4}\bar{g}_{\mu\nu}\left(  1-e^{-\chi\left(
\phi\right)  }\right)  ^{2} =0,\label{Eq Field Einstein 1 phi}
\end{equation}
and%
\begin{equation}
\bar{\square}\phi-\frac{\kappa_{\alpha}M_{P}}{\sqrt{2}}\frac{e^{-\chi\left(
\phi\right)  }\left(  1-e^{-\chi\left(  \phi\right)  }\right)  }%
{\sqrt{3\left(  1+be^{\chi\left(  \phi\right)  }\right)  }}%
=0.\label{Eq Field Einstein 2 phi}%
\end{equation}
The implicit dependence of $\chi\left(  \phi\right)  $ is obtained by
integrating (\ref{d chi / d psi}) which results in%
\begin{equation}
\phi\left(  \chi\right)  = M_{P}\sqrt{\frac{3}{2}}\left[  \chi+2\left(
\sqrt{1+be^{\chi}}-1\right) - 2\ln\left(  \frac{1+\sqrt{1+be^{\chi}}}{2}\right)  \right]
.\label{phi chi}%
\end{equation}
In the limit $b\rightarrow0$, we recover $\chi=\sqrt{\frac{2}{3}}\frac{\phi
}{M_{P}}.$

The expressions (\ref{Eq Field Einstein 1 phi}) and
(\ref{Eq Field Einstein 2 phi}) represent the final form of the field
equations in the Einstein frame considering that the nonlocal term contributes
as a small correction to the Starobinsky model.

\section{Inflation\label{sec - inflation}}

Let's start by computing the Friedmann equations. Considering the FLRW metric
in the form%
\[
ds^{2}=-dt^{2}+a^{2}\left(  t\right)  \left[  dx^{2}+dy^{2}+dz^{2}\right]  ,
\]
we obtain%
\begin{align}
H^{2} &  =\frac{1}{3M_{P}^{2}}\left[  \frac{1}{2}\dot{\phi}^{2}+V\left(
\phi\right)  \right]  ,\label{Friedmann 1}\\
\dot{H} &  =-\frac{\dot{\phi}^{2}}{2M_{P}^{2}},\label{Friedmann 2}%
\end{align}
and%
\begin{equation}
\ddot{\phi}+3H\dot{\phi}+V^{\prime}\left(  \phi\right)
=0,\label{Campo Phi em FRW}%
\end{equation}
where%
\begin{align}
V\left(  \phi\right)    & =\frac{\kappa_{\alpha}M_{P}^{2}}{4}\left(
1-e^{-\chi\left(  \phi\right)  }\right)  ^{2},\label{Potencial}\\
V^{\prime}\left(  \phi\right)    & =\frac{\kappa_{\alpha}M_{P}}{\sqrt{6}}%
\frac{e^{-\chi\left(  \phi\right)  }\left(  1-e^{-\chi\left(  \phi\right)
}\right)  }{\sqrt{1+be^{\chi\left(  \phi\right)  }}}.\label{DPotencial}%
\end{align}
The "prime" notation represents the derivative of the potential concerning
$\phi$. For consistency with the approximations performed, we will assume that
$\left\vert b\right\vert <0.1$. Besides, for negative $b$ we get an extra
constraint ($-e^{-\chi}<b<0$) due to the roots present in the Eq.
(\ref{phi chi}).

\begin{figure}[h]
   \centering
   \includegraphics[scale=0.55]{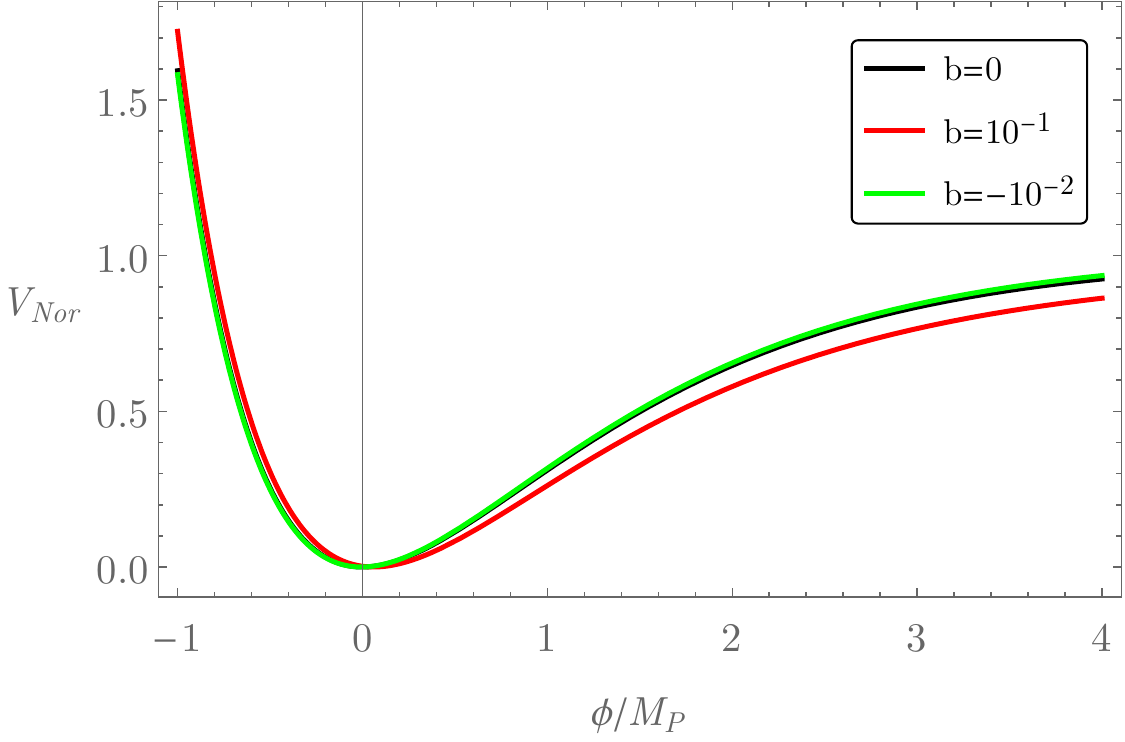}
   \caption{Plot of potential $V_{Nor}$ normalized by $\kappa_{\alpha
       }M_{P}^{2}/4$ as a function of $\phi/M_{p}$. The three curves were obtained
       with $b=0$ (black), $b=-10^{-2}$ (green) e $b=10^{-1}$ (red). The choice of
       $b=-10^{-2}$ comes from the fact that negative $b$ must respect the constraint
       $b>-e^{-\chi}$.}
   \label{fig1}
\end{figure}

The plot of the potential $V$ as a function of $\phi$ is shown in figure \ref{fig1}.

The most significant difference occurs for $b=10^{-1}$, but even in this case,
the curve behaves similarly to the Starobinsky potential. Thus, it is clear
that we have a slow-roll inflationary regime in the plateau region. Moreover,
for $b\neq0$, the minimum of the potential shifts to a value different from
the origin:%

\[
V^{\prime}\left(  \phi\right)  =0\Rightarrow\chi\left(  \phi_{\min}\right)
=0
\]
which, by the equation (\ref{phi chi}), results in%

\begin{equation}
\phi_{\min}=M_{P}\sqrt{6}\left[  \sqrt{1+b}-1-\ln\left(  \frac{1+\sqrt{1+b}%
}{2}\right)  \right]  .\label{Phi_min}%
\end{equation}
For the particular cases $b=10^{-1}$ and $b=-10^{-2}$, we obtain $\phi_{\min
}\approx0.0605$ $M_{P}$ and $\phi_{\min}\approx-0.0061$ $M_{P}$, respectively.
Despite this change, in the neighborhoods of $\phi_{\min}$, the potential
behaves like a quadratic potential $V\left(  \phi\right)  \sim\left(
\phi-\phi_{\min}\right)  ^{2}$. Therefore, at the end of the inflationary
regime, the period of coherent oscillations produces a cosmic dynamic
identical to the Starobinsky model, i.e. an effective equation of state
$\left\langle w\right\rangle \approx0$ and a period of expansion like a
matter-dominated universe \cite{MukBook}.

\subsection{Slow-roll regime}

The slow-roll inflationary regime occurs in the plateau region of the
potential where $\dot{\phi}^{2}\ll V\left(  \phi\right)  $. In this region,
the equations (\ref{Friedmann 1}), (\ref{Friedmann 2}) and
(\ref{Campo Phi em FRW}) can be approximated by%
\begin{equation}
H^{2}\approx\frac{V}{3M_{P}^{2}}\text{, \ }\frac{\dot{\phi}}{M_{P}}%
\approx-\frac{V^{\prime}}{\sqrt{3V}}\text{ \ and\ \ }\dot{H}\approx
-\frac{V^{\prime2}}{6V}.\label{Eq aproximadas}%
\end{equation}

Let's start by calculating the number of e-folds $N$ in slow-roll
leading-order. Using the Eqs. (\ref{Eq aproximadas}) and (\ref{d chi / d psi})
we get
\[
N \equiv\ln\left(  \frac{a_{end}}{a}\right)  \approx-\frac{1}{M_{P}^{2}}%
{\displaystyle\int\limits_{\phi}^{\phi_{end}}}
\frac{V\left(  \phi\right)  }{V^{\prime}\left(  \phi\right)  }d\phi
 \approx-\frac{3}{4}
{\displaystyle\int\limits_{\chi}^{\chi_{end}}}
\left[  be^{2\chi}+e^{\chi}\left(  1-b\right)  -1\right]  d\chi.
\]
Integrating this last expression and considering $\left\vert b\right\vert
<10^{-1}$ and $e^{\chi}\gg e^{\chi_{end}}$ we obtain%
\begin{equation}
N\approx\frac{3}{4}e^{\chi}\left(  1+\frac{b}{2}e^{\chi}\right)
.\label{N dominante}%
\end{equation}
By imposing the Starobinsky limit, the equation (\ref{N dominante}) can be
uniquely inverted. Thus,%
\begin{equation}
e^{\chi}=\frac{\sqrt{1+\frac{8}{3}bN}-1}{b}.\label{Chi_k}%
\end{equation}
Note that for $b<0$, we have an extra constraint given by $8bN>-3$. Hence, the
$b$ parameter is limited by the range%
\begin{equation}
-\frac{3}{8N}<b<0.1.\label{restricao b}%
\end{equation}
For a maximum of $60$ e-folds, we get $-0.00625<b<0.1$.

The next step is to compute the slow-roll parameters $\epsilon$ and $\eta$
defined as%
\[
\epsilon\equiv-\frac{\dot{H}}{H^{2}}\text{ \ \ and \ \ }\eta\equiv-\frac{1}%
{H}\frac{\dot{\epsilon}}{\epsilon}.
\]
In slow-roll leading-order, these parameters can be written in terms of the
potential and its derivatives:
\[
\text{ }\epsilon \approx\frac{M_{P}^{2}}{2}\left(  \frac{V^{\prime}\left(
\phi\right)  }{V\left(  \phi\right)  }\right)  ^{2}\text{ \ \ and \ \ }
\eta \approx2M_{p}^{\text{ }2}\left[  \frac{V^{\prime\prime}\left(
\phi\right)  }{V\left(  \phi\right)  }-\left(  \frac{V^{\prime}\left(
\phi\right)  }{V\left(  \phi\right)  }\right)  ^{2}\right].
\]
By carrying out the explicit calculations, we get%
\[
\epsilon  \approx\frac{4}{3}\left(  \frac{e^{-2\chi}}{1+be^{\chi}}\right)
\text{ \ \ and \ \ }
\eta  \approx-\frac{4}{3}e^{-\chi}\left(  \frac{2+3be^{\chi}}{\left(
1+be^{\chi}\right)  ^{2}}\right).
\]
Finally, substituting Eq. (\ref{Chi_k}) in these two expressions and
performing the suitable approximations, we obtain
\begin{equation}
\epsilon  \approx\frac{4}{3}\left[  \frac{b^{2}}{\left(  \sqrt{1+\frac{8}%
{3}bN}-1\right)  ^{2}\sqrt{1+\frac{8}{3}bN}}\right]\text{ \ and \ }
\eta  \approx-\frac{4b}{3}\left[  \frac{3\sqrt{1+\frac{8}{3}bN}-1}{\left(
\sqrt{1+\frac{8}{3}bN}-1\right)  \left(  1+\frac{8}{3}bN\right)  }\right].
\label{epsilon N}%
\end{equation}
Note that in the limit $b\rightarrow0$, we recover the results of the
Starobinsky model i.e.%
\[
\lim_{b\rightarrow0}\epsilon=\frac{3}{4}\frac{1}{N^{2}}\text{ \ \ and
\ \ }\lim_{b\rightarrow0}\eta=-\frac{2}{N}\text{.}%
\]

The equations (\ref{epsilon N}) ensure a slow-roll
inflationary regime, i.e. $\epsilon\ll1$ and $\eta\ll1$, whenever we have a
sufficiently large number of e-folds (e.g. $N\geq50$).\footnote{The $b$
parameter cannot be too close to the lower limit $-3/8N$.}

\subsection{Observational constraints\label{sec - constraints}}

Inflationary models can be constrained from observations of CMB anisotropies.
The constraint procedure is performed from the scalar and tensor power spectra
parameterized as \cite{Planck2013}%
\begin{equation}
\mathcal{P}_{s}=A_{s}\left(  \frac{k}{k_{\ast}}\right)  ^{n_{s}}\text{ \ \ and
\ \ \ }\mathcal{P}_{t}=A_{t}\left(  \frac{k}{k_{\ast}}\right)  ^{n_{t}%
},\label{Paremtrizacao dos PSs}%
\end{equation}
where $A_{s}$ and $A_{t}$ are the scalar and tensor amplitudes, $n_{s}$ and
$n_{t}$ are the scalar and tensor spectral indices, and $k_{\ast}$ is a
reference scale (pivot scale). It is also usual to define the tensor-to-scalar
ratio%
\begin{equation}
r\equiv\frac{A_{t}}{A_{s}}.\label{tensor to scalar ratio}%
\end{equation}
Moreover, for inflationary models of a single canonical scalar field (such as
the proposed model in its approximate form), the consistency relation
$n_{t}=-r/8$ is always verified. Thus, there are only three free parameters
that can be represented by $A_{s}$, $n_{s}$ and $r$.

In slow-roll leading-order, we know that \cite{Baumann2018,SilSoMe2021}%
\begin{equation}
n_{s}=1+\eta-2\epsilon\text{ \ \ and \ \ }r=16\epsilon. \label{ns e r}%
\end{equation}
From Eqs. (\ref{epsilon N}) we see that $n_{s}$ and $r$
depend on the number of e-folds $N$ and the parameter $b$.

\begin{figure}[h]
   \centering
   \includegraphics[scale=0.44]{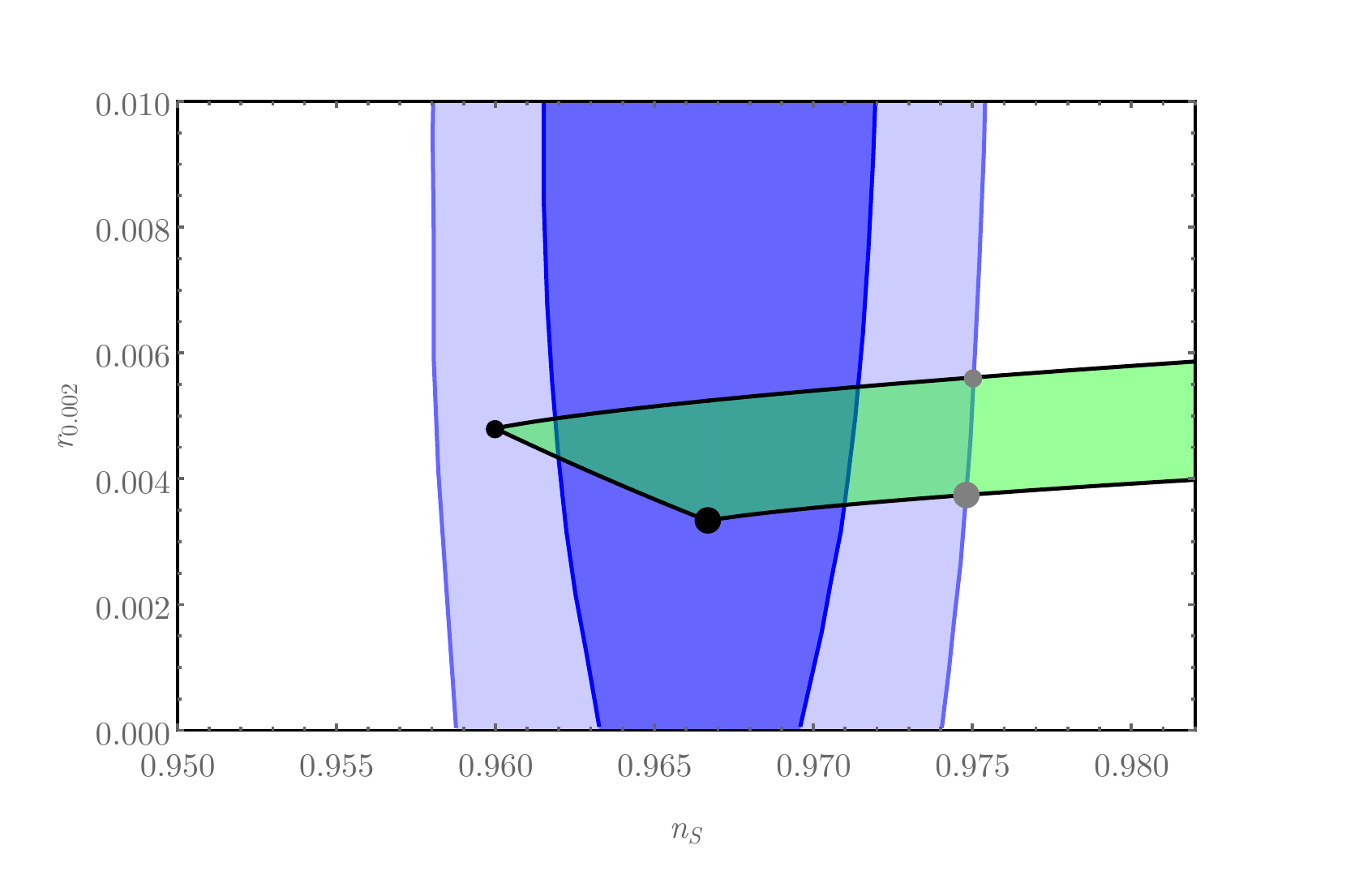}
      \caption{Parameter space $n_{s}\times r$ which include the
       observational constraints $68\%$ (dark blue) and $95\%$ (light blue) C.L.
       \cite{Bicep2021} and the theoretical evolution of the model (green) calculated
       from Eq. (\ref{ns e r}). The constraint is made considering $b<0$ and $50\leq
       N\leq60$. The black circles represent the Starobinsky model ($b=0$) for $N=50$
       (smaller one) and $N=60$ (bigger one). As $\left\vert b\right\vert $ increases
       the curves move to the right (light green region) increasing the
       tensor-to-scalar ratio and the scalar tilt values. The grey circles take into
       account the maximum values of $\left\vert b\right\vert $ still consistent with
       the region of $95\%$ C.L.. In this case, $N=50$ and $N=60$ correspond to
       $b=-0.0060$ and $b=-0.0047$, respectively.}
   \label{fig2}
\end{figure}

\begin{figure}[h]
   \centering
   \includegraphics[scale=0.47]{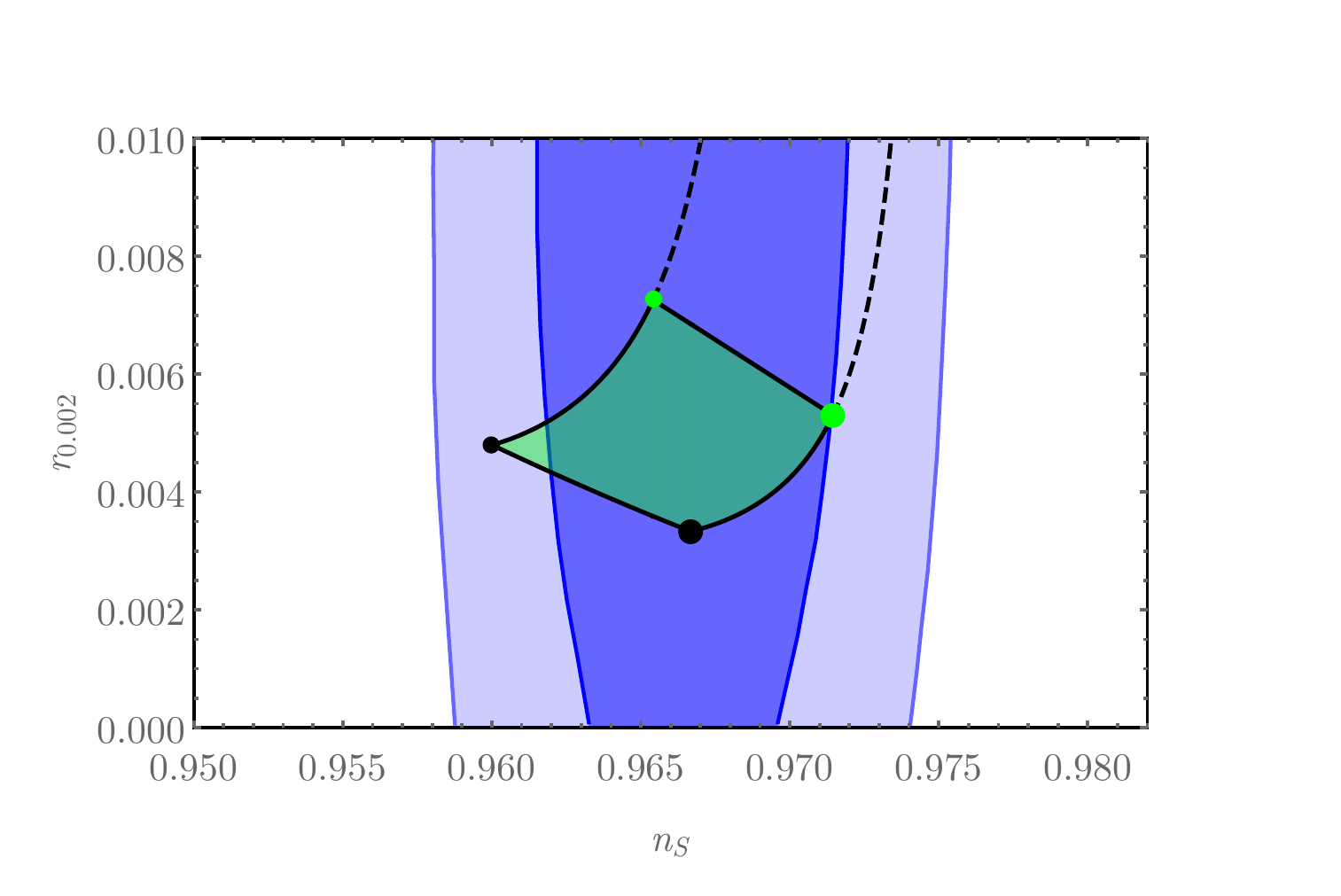}
      \caption{Similar analyzes as figure \ref{fig2} but now considering $b>0$. The
       green circles correspond to the upper bound $b=10^{-1}$ where the perturbative
       approach is still valid. In this case, $N=50$ and $N=60$ correspond to
       $\left(  n_{s},r\right)  =\left(  0.9654,0.0073\right)  $ and $\left(
       n_{s},r\right)  =\left(  0.9714,0.0053\right)  $, respectively. The dotted
       lines extrapolate Eq. (\ref{ns e r}) to $b>10^{-1}$.}
   \label{fig3}
\end{figure}

The comparison with the observations through the parameter space $n_{s}\times
r$ must be performed by separating the cases of $b$ positive and $b$ negative.
Figures \ref{fig2} and \ref{fig3} show the cases $b<0$ and $b>0$, respectively.%

Figure \ref{fig2} shows that for negative $b$, the observational data constrain in a
very restrictive way the value of $b$. Within $50\leq N\leq60$, we get
$-0.006\leq b<0$. In addition, the variation of $b$ has little effect on the
value of the tensor-to-scalar ratio. For $N=50$ and $N=60$ within $95\%$ C.L.,
we obtain $0.0048<r<0.0056$ and $0.0033<r<0.0037$, respectively.

On the other hand, from figure \ref{fig3}, we see that the value of positive $b$ is
little constrained by the observations. The entire region encompassing
$0<b\leq10^{-1}$ and $50\leq N\leq60$ is within the range of observationally
values allowed by $95\%$ C.L.. Moreover, we realize that $b$ positive admits a
more significant variation of the tensor-to-scalar ratio than $b$ negative.

In addition to the restrictions on parameters $b$ and $N$, we can constrain
the parameter $\kappa_{\alpha}$ from the observation of the scalar amplitude
$A_{s}$. In slow-roll leading-order, the scalar amplitude can be written as
\cite{Baumann2018,SilSoMe2021}%
\begin{equation}
A_{s}=\frac{1}{12\pi^{2}M_{p}^{6}}\frac{V^{3}}{V^{\prime2}}. \label{As}%
\end{equation}
Substituting Eqs. (\ref{Potencial}), (\ref{DPotencial}) and (\ref{Chi_k}) in the last expression,
we get%
\begin{equation}
\kappa_{\alpha}=\frac{2^{7}\pi^{2}A_{s}M_{p}^{2}b^{2}}{\left(  \sqrt
{1+\frac{8}{3}bN}-1\right)  ^{2}\sqrt{1+\frac{8}{3}bN}}. \label{kappa N}%
\end{equation}
Using Eqs. (\ref{epsilon N}) and (\ref{ns e r}) and the value $A_{s}%
=2.1\times10^{-9}$ \cite{Planck2018CosPar}, we can rewrite $\kappa_{\alpha}$
as
\begin{equation}
\kappa_{\alpha}=6\pi^{2}rA_{s}M_{p}^{2}=1.25\times10^{-7}rM_{p}^{2}.
\label{kappa r}%
\end{equation}
Therefore, taking into account that the maximum variation of the
tensor-to-scalar ratio is $0.0033\leq r\leq0.0073$, we obtain%

\begin{equation}
4\times10^{-10}M_{p}^{2}\leq\kappa_{\alpha}\leq9\times10^{-10}M_{p}^{2}.
\label{range kappa alpha}%
\end{equation}
The last result confirms that the inflation energy scale $E_{\inf}\sim
\kappa_{\alpha}^{1/2}\sim10^{-5}$ $M_{p}$.

\section{Final Comments\label{sec - final comments}}

In this work, we investigate how the inclusion of a nonlocal term $R\ln\left(
\square\right)  R$ changes the Starobinsky inflation. We consider that this
term provides a small correction to the Starobinsky model. Using a
perturbative approach, we show that the field equations reduce to local
equations, which in the Einstein frame can be described by a canonical scalar
field minimally coupled to general relativity. The $b$ parameter, which
measures the effectiveness of the nonlocal term concerning the Starobinsky
term, was constrained in Section \ref{sec - constraints}. For negative $b$ we
obtained $\left\vert b\right\vert <0.006$, and for positive $b$ we did not
obtain any constraint within the perturbative context. The results achieved in
Section \ref{sec - constraints} are similar to those presented in Refs.
\cite{CaSoShaSta2018,SilMe2022}, where the authors study the influence
of the $R\square R$ term on Starobinsky inflation. This similarity
shows that in the context of small corrections, the contribution of the
nonlocal term occurs essentially through the first term of the series in Eq.
(\ref{rep. serie 2}).

The proposed model is inserted in a context of effective theories whose
validity energy range is below the Planck scale and far above the masses
of the matter fields. In this context, the $b$ parameter is not a free
parameter but depends on the quantity of matter scalar fields present in the
original theory. If we take into account only the Higgs field and fix the
renormalization point on the inflationary energy scale ($\mu^{2}\sim\kappa
_{0}$), we get, from Eqs. (\ref{alpha definition}) and
(\ref{range kappa alpha}),%
\[
\left\vert b\right\vert =\frac{4\kappa_{0}}{M_{P}^{2}}\frac{20\left(
6\xi-1\right)  ^{2}}{11520\pi^{2}}\sim10^{-13}\left(  6\xi-1\right)  ^{2},
\]
where $\kappa_{0}\simeq\kappa_{\alpha}\simeq5\times10^{-10}M_{p}^{2}$. This
equation shows that for $\left\vert b\right\vert \sim10^{-3}$ (see figure
\ref{fig2}), it is necessary $\xi\sim2\times10^{4}$. The high value of the
non-minimum coupling constant $\xi$ is consistent with the Higgs inflation
model proposed in \cite{BezShapo2008}\textbf{ }where $\xi\sim5\times
10^{4}\sqrt{\lambda}$.\footnote{The $\lambda$ parameter is the quartic Higgs
field self-coupling \cite{Steinwachs2019}\textbf{.}} It is also worth
mentioning that the inclusion of new scalar degrees of freedom increases the
value of $N_{s}$ present in Eq. (\ref{alpha definition}) and causes $\xi$ to
decrease to a fixed value of $b$.

The discussion in the previous paragraph supports the idea that two different
approaches can be used to deal with light matter fields in an inflationary context
of modified gravity. The first approach considers these fields
explicitly and analyzes how they affect inflation (see, for example, ref.
\cite{GunSte2019}). The second one uses the idea of effective theories, which
treat all matter fields collectively and transfer their effects to the
gravitational degrees of freedom. In principle, the second approach is only
correct if the second nonlocal term $\left(  2\beta
/M_{P}^{2}\right)  C_{\kappa\rho\alpha\beta}\ln\left(  \square\right)
C^{\kappa\rho\alpha\beta}$ is also included. Nevertheless, only considering matter fields of
the standard model (minimalist model), the value of $c\equiv$ $4\beta
\kappa_{0}/M_{P}^{2}$, which measures the effectiveness of the second nonlocal
term concerning the Starobinsky term, is extremely small ($\left\vert
c\right\vert \sim10^{-8}$) \cite{DonMen2014}. Thus, unless the matter degrees
of freedom increase by several orders of magnitude,\footnote{Another
possibility would be to include some non-minimal coupling between vector or
fermionic fields with gravitation.} it is reasonable to expect that the term
$C_{\kappa\rho\alpha\beta}\ln\left(  \square\right)  C^{\kappa\rho\alpha\beta
}$ will be negligible in the inflationary context.

The conclusion presented above disregards the inclusion of new
degrees of freedom arising from the nonlocal term. It occurs because the
approximation of small corrections performed in section \ref{sec - perturbative} transfers any
extra degrees of freedom to the scalaron associated with the $R^{2}$ term. In
this sense, possible issues generated by new degrees of freedom are being
neglected. For example, it is common for degrees of freedom engendered by
higher-order models to present ghost-like fields. Even so, the possible
pathologies associated with these fields do not necessarily influence the
inflationary dynamics. A situation where this statement is true occurs in the
local model $\mathcal{L}=R+aR^{2}+bR\square R$.\footnote{This model can be
obtained by considering only the first two terms of the series expansion of
$\ln\left(  \square\right)  $. See Eq. (\ref{App}).} In this model, the term $R\square
R$ contributes an extra ghost field, but the perturbations of this field are
negligible in the inflationary context \cite{SilMe2022}. It is also worth mentioning that
the results found in Ref. \cite{SilMe2022} are similar to those obtained in Ref. \cite{CaSoShaSta2018}
where the authors consider only small corrections of $R\square R$ to the
Starobinsky model. Due to its nonlocal feature, the study of extra degrees of
freedom generated by the term $R\ln\left(  \square\right)  R$ is more complex than that performed in local models.
This study will be carried out in future works.

\acknowledgments

J. Bezerra-Sobrinho thanks PIBIC CNPq/UFRN-RN (Brazil) for financial support and L. G.
Medeiros acknowledges CNPq-Brazil (Grant No. 308380/2019-3) for partial financial support.

\appendix

\section{Jordan Frame\label{ApJD}}

In order to rewrite the gravitational action (\ref{acao S}) in the Jordan
frame, we start by defining two parameters%
\[
\lambda_{1}=R\text{ \ \ \ and \ \ \ }\lambda_{2}=\ln\left(  \frac{\square}{\mu^{2}%
}\right)  R\text{.}%
\]
From these parameters, we build a new action in the form%
\[
\bar{S}  = \frac{M_{P}^{2}}{2}\int d^{4}x\sqrt{-g}\left\{  \lambda_{1}%
+\frac{1}{2\kappa_{0}}\lambda_{1}^{2}+\frac{2\alpha}{M_{P}^{2}}\lambda
_{1}\lambda_{2}  \theta_{1}\left(  R-\lambda_{1}\right)  +\theta_{2}\left[
\ln\left(  \frac{\square}{\mu^{2}}\right)  \lambda_{1}-\lambda_{2}\right]
\right\},
\]
where the fields $\theta_{1}$ and $\theta_{2}$ are Lagrange multipliers. By
taking the variation of $\bar{S}$ concerning $\theta_{1}$ and $\theta_{2}$ and
using the field equations, we easily realize that $\bar{S}$ and the original
action are equivalent on-shell.

The next step is to compute the variation of $\bar{S}$ with respect to
$\lambda_{1}$ and $\lambda_{2}$. To perform this calculation, we will use the
series representation (\ref{rep. serie 2}). Thus,%
\[
\int d^{4}x\sqrt{-g}\theta_{2}\ln\left(  \frac{\square}{\mu^{2}}\right)
\lambda_{1} = \sum_{n=1}^{\infty}\sum_{k=0}^{n}\frac{\left(  -1\right)
^{n-1}}{\mu^{2k}n}\frac{\left(  -1\right)  ^{n-k}n!}{k!\left(  n-k\right)
!} \int d^{4}x\sqrt{-g}\theta_{2}\square^{k}\lambda_{1}.
\]
Applying Leibniz rule $k$ times in $\theta_{2}\square^{k}\lambda_{1}$ and
neglecting the surface terms we get%
\begin{equation}
S = \frac{M_{P}^{2}}{2}\int d^{4}x\sqrt{-g}\left\{  \lambda_{1}+\frac
{1}{2\kappa_{0}}\lambda_{1}^{2}+\frac{2\alpha}{M_{P}^{2}}\lambda_{1}%
\lambda_{2} +\theta_{1}\left(  R-\lambda_{1}\right)  +\lambda_{1}\ln\left(
\frac{\square}{\mu^{2}}\right)  \theta_{2}-\theta_{2}\lambda_{2}\right\}
.\label{S bar 1}%
\end{equation}
Thereby, the variations concerning $\lambda_{1}$ and $\lambda_{2}$ of the
above expression result in the field equations
\begin{align*}
1+\frac{1}{\kappa_{0}}\lambda_{1}+\frac{2\alpha}{M_{P}^{2}}\lambda_{2}%
-\theta_{1}+\ln\left(  \frac{\square}{\mu^{2}}\right)  \theta_{2} &  =0,\\
\frac{2\alpha}{M_{P}^{2}}\lambda_{1}-\theta_{2} &  =0.
\end{align*}
By inverting the last equations for $\lambda_{1}$ and $\lambda_{2}$ and
substituting the result in Eq. (\ref{S bar 1}) we obtain%
\begin{equation}
S  = \frac{M_{P}^{2}}{2}\int d^{4}x\sqrt{-g}\left\{  \theta_{1}R+\frac
{M_{P}^{2}}{2\alpha}\left(  1-\theta_{1}\right)  \theta_{2}
+  \frac{1}{2\kappa_{0}}\left(  \frac{M_{P}^{2}}{2\alpha}\right)
^{2}\theta_{2}^{2}+\frac{M_{P}^{2}}{2\alpha}\theta_{2}\ln\left(  \frac
{\square}{\mu^{2}}\right)  \theta_{2}\right\}.\label{acao frame Jordan}%
\end{equation}
Finally, we define
\[
\theta=\theta_{1}\text{, \ \ \ }\theta_{2}=\frac{2\alpha\kappa_{0}}{M_{P}^{2}%
}\lambda\text{ \ \ \ and \ \ \ }b=\frac{4\alpha\kappa_{0}}{M_{P}^{2}},
\]
and we achieve the equation (\ref{acao frame Jordan lambda}), which represents the original action in the Jordan frame.


\section{Metric equation in Jordan frame\label{Ap 3}}

We start by taking the variation $\delta_{g}$ in the action
(\ref{acao em partes}) concerning $g^{\mu\nu}$:%
\begin{equation}
\delta_{g}S=\frac{M_{P}^{2}}{2}\left\{  \int d^{4}x\sqrt{-g}\left[
\theta\delta_{g}R-\frac{1}{2}\left[  \theta R+\kappa_{0}\left(  1-\theta
\right)  \lambda+\frac{\kappa_{0}}{2}\lambda^{2}\right]  g_{\mu\nu}\delta
g^{\mu\nu}\right]  +\delta_{g}S_{NL}\right\}  ,\label{Variation metric}%
\end{equation}
where \cite{CapLau2011}
\begin{equation}
\int d^{4}x\sqrt{-g}\theta\delta_{g}R=\int d^{4}x\sqrt{-g}\left[  \theta
R_{\mu\nu}+\left(  \square\theta\right)  g_{\mu\nu}-\left(  \nabla_{\mu}%
\nabla_{\nu}\theta\right)  \right]  \delta_{g}g^{\mu\nu}.\label{termo 1}%
\end{equation}
Let's compute
\begin{equation}
\delta_{g}S_{NL}=\frac{M_{P}^{2}\kappa_{0}b}{4}\sum_{n=1}^{\infty}\sum
_{k=0}^{n}\frac{\left(  -1\right)  ^{n-1}}{n}\frac{\left(  -1\right)
^{n-k}n!}{k!\left(  n-k\right)  !}\frac{1}{\mu^{2k}}\delta_{g}\int d^{4}%
x\sqrt{-g}\lambda\left(  \square^{k}\lambda\right)
.\label{Variacao em g nao local}%
\end{equation}
The first step is to expand $\square^{k}$ as
\[
\square^{k}=\nabla_{\nu_{1}}\nabla^{\nu_{1}}\nabla_{\nu_{2}}\nabla^{\nu_{2}%
}...\nabla_{\nu_{k-1}}\nabla^{\nu_{k-1}}\nabla_{\nu_{k}}\nabla^{\nu_{k}}.
\]
Thus,%
\begin{align*}
\delta_{g}\int d^{4}x\sqrt{-g}\lambda\left(  \square^{k}\lambda\right)   &
=\delta_{g}\int d^{4}x\sqrt{-g}\lambda\nabla_{\nu_{1}}\nabla^{\nu_{1}}%
\nabla_{\nu_{2}}\nabla^{\nu_{2}}...\nabla_{\nu_{k-1}}\nabla^{\nu_{k-1}}%
\nabla_{\nu_{k}}\nabla^{\nu_{k}}\lambda\\
&  =\int d^{4}x\left(  \delta_{g}\sqrt{-g}\right)  \lambda\left(  \square
^{k}\lambda\right)  \\
&  +\int d^{4}x\sqrt{-g}\lambda\left[  \delta_{g}\left(  \nabla_{\nu_{1}%
}\nabla^{\nu_{1}}\right)  \right]  \nabla_{\nu_{2}}\nabla^{\nu_{2}}%
...\nabla_{\nu_{k-1}}\nabla^{\nu_{k-1}}\nabla_{\nu_{k}}\nabla^{\nu_{k}}%
\lambda\\
&  +\int d^{4}x\sqrt{-g}\lambda\nabla_{\nu_{1}}\nabla^{\nu_{1}}\left[
\delta_{g}\left(  \nabla_{\nu_{2}}\nabla^{\nu_{2}}\right)  \right]
...\nabla_{\nu_{k-1}}\nabla^{\nu_{k-1}}\nabla_{\nu_{k}}\nabla^{\nu_{k}}%
\lambda+...\\
&  +\int d^{4}x\sqrt{-g}\lambda\nabla_{\nu_{1}}\nabla^{\nu_{1}}\nabla_{\nu
_{2}}\nabla^{\nu_{2}}...\left[  \delta_{g}\left(  \nabla_{\nu_{k-1}}%
\nabla^{\nu_{k-1}}\right)  \right]  \nabla_{\nu_{k}}\nabla^{\nu_{k}}\lambda\\
&  +\int d^{4}x\sqrt{-g}\lambda\nabla_{\nu_{1}}\nabla^{\nu_{1}}\nabla_{\nu
_{2}}\nabla^{\nu_{2}}...\nabla_{\nu_{k-1}}\nabla^{\nu_{k-1}}\left[  \delta
_{g}\left(  \nabla_{\nu_{k}}\nabla^{\nu_{k}}\right)  \lambda\right]  .
\end{align*}
Integrating by parts several times, we get
\begin{align*}
\delta_{g}\int d^{4}x\sqrt{-g}\lambda\left(  \square^{k}\lambda\right)   &
=\int d^{4}x\left(  \delta_{g}\sqrt{-g}\right)  \lambda\left(  \square
^{k}\lambda\right)  +\int d^{4}x\sqrt{-g}\lambda\left[  \delta_{g}\left(
\nabla_{\nu_{1}}\nabla^{\nu_{1}}\right)  \right]  \square^{k-1}\lambda\\
&  +\int d^{4}x\sqrt{-g}\left[  \square\lambda\right]  \left[  \delta
_{g}\left(  \nabla_{\nu_{2}}\nabla^{\nu_{2}}\right)  \right]  \square
^{k-2}\lambda+...\\
&  +\int d^{4}x\sqrt{-g}\left[  \square^{k-2}\lambda\right]  \left[
\delta_{g}\left(  \nabla_{\nu_{k-1}}\nabla^{\nu_{k-1}}\right)  \right]
\square\lambda\\
&  +\int d^{4}x\sqrt{-g}\left[  \square^{k-1}\lambda\right]  \left[
\delta_{g}\left(  \nabla_{\nu_{k}}\nabla^{\nu_{k}}\right)  \right]  \lambda.
\end{align*}
In compact notation, the above expression can be written as
\[
\delta_{g}\int d^{4}x\sqrt{-g}\lambda\left(  \square^{k}\lambda\right)  =\int
d^{4}x\left(  \delta_{g}\sqrt{-g}\right)  \lambda\left(  \square^{k}%
\lambda\right)  +%
{\displaystyle\sum\limits_{l=1}^{k}}
I_{l,k},
\]
where%
\[
I_{l,k}=\int d^{4}x\sqrt{-g}\square^{l-1}\lambda\left[  \delta_{g}\left(
\nabla_{\nu}\nabla^{\nu}\right)  \square^{k-l}\lambda\right]  .
\]
The next step is working with the integral $I_{l,k}$. Using the relation
$\delta\sqrt{-g}=-\frac{1}{2}\sqrt{-g}g_{\mu\nu}\delta g^{\mu\nu}$, we obtain%
\begin{align*}
I_{l,k} &  =\int d^{4}x\sqrt{-g}\square^{l-1}\lambda\left[  \delta_{g}\left(
\frac{1}{\sqrt{-g}}\partial_{\mu}\left(  \sqrt{-g}g^{\mu\nu}\partial
_{v}\right)  \right)  \square^{k-l}\lambda\right]  \\
&  =\frac{1}{2}\int d^{4}x\sqrt{-g}\left(  \square^{l-1}\lambda\right)
\left(  \square^{k-l+1}\lambda\right)  g_{\mu\nu}\delta_{g}g^{\mu\nu}\\
&  +\frac{1}{2}\int d^{4}x\sqrt{-g}\left[  \nabla_{\rho}\square^{l-1}%
\lambda\right]  \left[  \nabla^{\rho}\square^{k-l}\lambda\right]  g_{\mu\nu
}\delta_{g}g^{\mu\nu}\\
&  -\int d^{4}x\sqrt{-g}\left[  \nabla_{\mu}\square^{l-1}\lambda\right]
\left[  \nabla_{v}\square^{k-l}\lambda\right]  \delta_{g}g^{\mu\nu}.
\end{align*}
Thus,%
\begin{align*}
\delta_{g}\int d^{4}x\sqrt{-g}\lambda\left(  \square^{k}\lambda\right)   &
=-\frac{1}{2}\int d^{4}x\sqrt{-g}\lambda\left(  \square^{k}\lambda\right)
g_{\mu\nu}\delta g^{\mu\nu}\\
&  +
{\displaystyle\sum\limits_{l=1}^{k}}
\frac{1}{2}\int d^{4}x\sqrt{-g}\left(  \square^{l-1}\lambda\right)  \left(
\square^{k-l+1}\lambda\right)  g_{\mu\nu}\delta g^{\mu\nu}\\
&  +{\displaystyle\sum\limits_{l=1}^{k}}
\frac{1}{2}\int d^{4}x\sqrt{-g}\left[  \nabla_{\rho}\square^{l-1}%
\lambda\right]  \left[  \nabla^{\rho}\square^{k-l}\lambda\right]  g_{\mu\nu
}\delta g^{\mu\nu}\\
&  -{\displaystyle\sum\limits_{l=1}^{k}}
\int d^{4}x\sqrt{-g}\left[  \nabla_{\mu}\square^{l-1}\lambda\right]  \left[
\nabla_{v}\square^{k-l}\lambda\right]  \delta_{g}g^{\mu\nu}.
\end{align*}
By substituting this last result in Eq. (\ref{Variacao em g nao local}), we get%
\begin{align}
\delta_{g}S_{NL} &  =-\frac{b\kappa_{0}}{4}\int d^{4}x\sqrt{-g}\delta
g^{\mu\nu}g_{\mu\nu}\lambda\ln\left(  \frac{\square}{\mu^{2}}\right)
\lambda\nonumber\\
&  +\frac{b\kappa_{0}}{4}\int d^{4}x\sqrt{-g}\delta g^{\mu\nu}g_{\mu\nu}%
\sum_{n=1}^{\infty}\sum_{k=1}^{n}\frac{\left(  -1\right)  ^{n-1}}{n}%
\frac{\left(  -1\right)  ^{n-k}n!}{k!\left(  n-k\right)  !}\frac{1}{\mu^{2k}}%
{\displaystyle\sum\limits_{l=1}^{k}}
\left(  \square^{l-1}\lambda\right)  \left(  \square^{k-l+1}\lambda\right)
\nonumber\\
&  +\frac{b\kappa_{0}}{4}\int d^{4}x\sqrt{-g}\delta g^{\mu\nu}g_{\mu\nu}%
\sum_{n=1}^{\infty}\sum_{k=1}^{n}\frac{\left(  -1\right)  ^{n-1}}{n}%
\frac{\left(  -1\right)  ^{n-k}n!}{k!\left(  n-k\right)  !}\frac{1}{\mu^{2k}}%
{\displaystyle\sum\limits_{l=1}^{k}}
\left[  \nabla_{\rho}\square^{l-1}\lambda\right]  \left[  \nabla^{\rho}%
\square^{k-l}\lambda\right]  \nonumber\\
&  -\frac{b\kappa_{0}}{2}\int d^{4}x\sqrt{-g}\delta_{g}g^{\mu\nu}\sum
_{n=1}^{\infty}\sum_{k=1}^{n}\frac{\left(  -1\right)  ^{n-1}}{n}\frac{\left(
-1\right)  ^{n-k}n!}{k!\left(  n-k\right)  !}\frac{1}{\mu^{2k}}%
{\displaystyle\sum\limits_{l=1}^{k}}
\left[  \nabla_{\mu}\square^{l-1}\lambda\right]  \left[  \nabla_{v}%
\square^{k-l}\lambda\right]  .\label{Variacao g SNL}%
\end{align}
Finally, we substitute Eqs. (\ref{termo 1}) and (\ref{Variacao g SNL}) in Eq.
(\ref{Variation metric}), and we achieve the metric equation (\ref{Field1}).

\end{document}